# Video Microscopy of Colloidal Suspensions and Colloidal Crystals

Piotr Habdas and Eric R. Weeks*
*Physics Department, Emory University, Atlanta, GA 30322, USA*
*Phone: 404-727-4479, Fax: 404-727-0873*
* Corresponding author.  Email: weeks@physics.emory.edu

**Abstract**

Colloidal suspensions are simple model systems for the study of phase transitions.  Video microscopy is capable of directly imaging the structure and dynamics of colloidal suspensions in different phases.  Recent results related to crystallization, glasses, and 2D systems complement and extend previous theoretical and experimental studies.  Moreover, new techniques allow the details of interactions between individual colloidal particles to be carefully measured.  Understanding these details will be crucial for designing novel colloidal phases and new materials, and for manipulating colloidal suspensions for industrial uses.

## 1. Introduction

Video optical microscopy has long been the tool of biologists, used to examine the structure and function of cells [1*].  In the past decade, physicists and chemists have begun using optical microscopy techniques to study the structure and dynamics of colloidal suspensions.  Microscopy provides useful complementary information to other techniques such as light scattering or neutron scattering.  The applications have included studying the phase behavior of colloidal suspensions, and learning details about the fundamental interactions between colloidal particles.  This section of the paper will briefly describe microscopy techniques useful for studying colloids, and the remainder of the paper will discuss recent experiments which have utilized microscopy in a variety of colloidal systems.

Traditional optical microscopy is comprised of a large variety of specialized techniques [1*].  When applied to colloidal suspensions, the primary requirement is that the index of refraction of the colloidal particles must be closely matched to that of the solvent; otherwise, scattering prevents the microscope from looking deep inside the sample.  Several microscopy techniques are in fact optimized for looking at nearly-index-matched samples [1*].  The index-matching requirement is relaxed if the particle concentration is low. Another possibility is to study dense colloidal suspensions in thin cells, where only one or two layers of particles need to be viewed. Besides being experimentally tractable, such quasi-two-dimensional systems are also of theoretical interest (Sec. 2.3). Also, some specialized techniques [1*,2,3], such as total internal reflection microscopy [2], provide high-resolution information about the colloidal particles close to the surface of the sample chamber.

One key technique taken from biology is to use fluorescent markers, to help view particles which are otherwise optically transparent [1*]. Colloidal particles can be purchased stained with fluorescent dyes, from companies such as Molecular Probes, Interfacial Dynamics Corporation, or Bangs Laboratories. Fluorescence microscopy is a powerful technique, as different dyes can be used to color different components of a complex system that are then distinguished with optical filters. Moreover, fluorescent colloidal particles can be studied using the powerful technique of laser scanned confocal microscopy [4*,5*,6,7*,8,9]. Confocal microscopy uses an optical microscope to focus a laser onto the sample, where it excites the fluorescence in the dyed particles. The laser scans across the sample in $x$ and $y$, and the emitted fluorescent light is de-scanned and focused onto a detector. Before entering the detector, the light from the focal point of the microscope objective is refocused onto a screen with a pinhole, which acts as a spatial filter, rejecting out of focus light and restricting the depth of focus of the image. Confocal microscopes thus get clean two-dimensional images from deep within samples, such as shown in Fig. 1 and 2. By scanning at different depths, three-dimensional pictures can be obtained, as shown in Fig. 3.

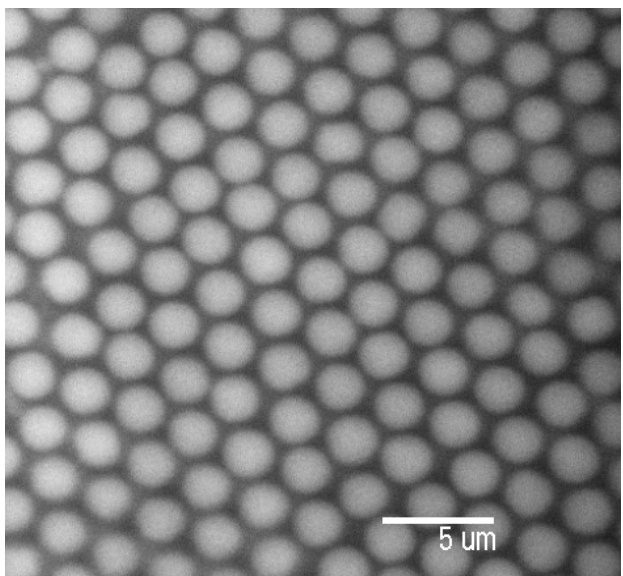

Fig.1. A snapshot of a hexagonal close packed crystal layer of 2.2 μm diameter fluorescent PMMA particles obtained using a confocal microscope with a 100× objective.

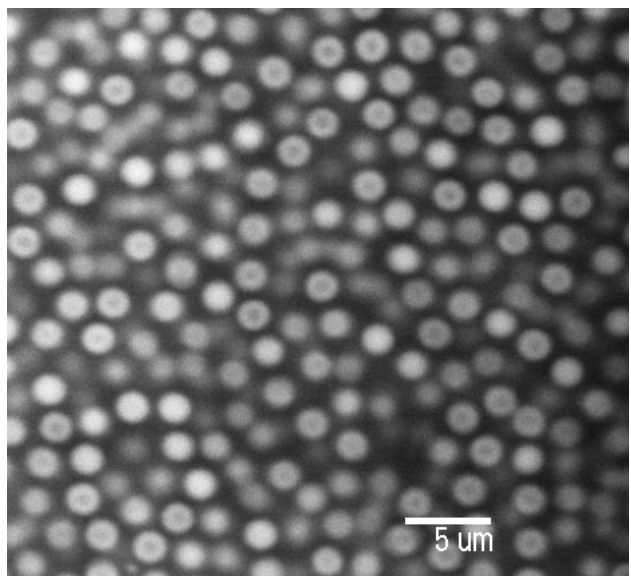

Fig.2. A snapshot of a fluid of 2.2 μm diameter fluorescent PMMA particles obtained using a confocal microscope with a 100× objective.

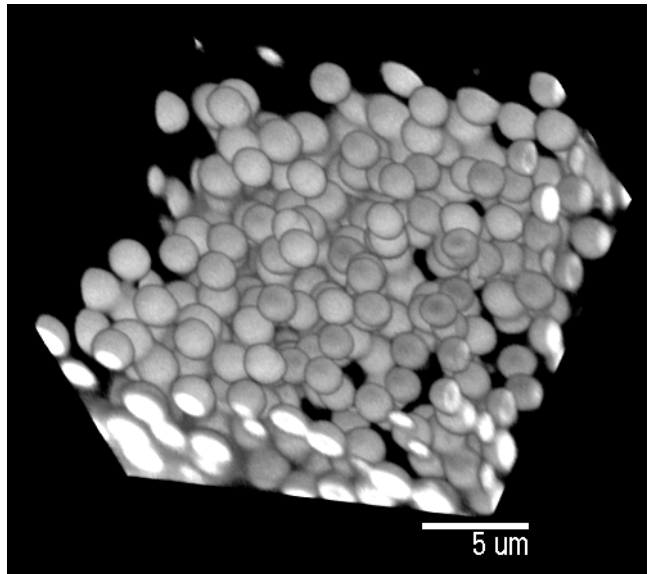

Fig.3. A three-dimensional image of 2.2 μm diameter fluorescent PMMA colloidal particles in a colloidal gel phase obtained using confocal microscope with a 100× objective.

Image analysis techniques alone can provide some information about colloidal systems, and certainly images of colloidal crystals are useful for learning about structural arrangements. While not exclusively a microscopy technique, particle tracking – tracking the motion of individual colloidal particles [10*] – provides useful information about the dynamics of colloidal phases (Sec. 2) and the interactions and forces between colloidal particles (Sec. 3). From images such as those shown in Figs. 1 and 2, the centers of particles can be found by computing the brightness-weighted centroid of each particle in the image. Using this method, the particle positions are determined to a precision of as good as 10-30 nm, limited by the number of pixels per particle and the noise in the original image, rather than the optical resolution of the microscope (typically 200 nm for a 100× objective). Moreover, computer techniques can automate the tracking of several hundred particles simultaneously [10*].

Another important technique is that of laser tweezers [11,12]. Laser tweezers use a focused laser beam to trap a small dielectric particle. More powerful techniques utilize various ways to create multiple traps, or traps that are spatially extended in one or more dimensions, see for example [13*, 14] and other references in Sec. 3. These techniques allow experimentalists to seize individual particles and move them to new locations [15, 16], or to confine particles to a small region to study their interactions [14]. Details of some of these experiments are discussed in Sec. 2 and Sec. 3.

## 2. Imaging colloidal phases

Colloids have been used as model systems to study phase transitions since 1986, when Pusey and van Megen showed that certain colloids follow the phase behavior predicted for hard spheres, and are thus simple models of the packing of atoms [17*]. In particular, they found that the spatial arrangement of colloids can mimic liquids, crystals, and glasses (see Figs. 1 and 2). The large size of colloidal particles makes them particularly easy to observe directly using video microscopy, and thus it is not surprising that many experimentalists have carefully studied the real-space structure of the different colloidal phases, and the local motion of the colloidal particles near the phase transitions.

*2.1 Crystals*

The most visually pleasing colloidal phase is the crystalline phase [18,19]. To the naked eye, colloidal crystals appear iridescent, as they Bragg scatter visible light; this is the reason opals are iridescent, for example. Under the microscope, the ordering of a colloidal crystal is also visually striking, as seen in Fig. 1 for example. Entropy causes the spontaneous formation of colloidal crystals in dense colloidal suspensions. For example, in some simple cases colloidal particles can be considered as hard spheres: the free volume for hard spheres in a hexagonal close packed (HCP) or face centered cubic (FCC) crystal lattice (maximum packing volume fraction of 74%) is greater than that of disordered spheres (maximum packing $\approx$ 64%), and so the entropy is higher in the crystalline state due to the increased local vibrational degrees of freedom of the particles. Thus many colloidal crystals are "random hexagonal close-packed" (RHCP), hexagonal layers that are randomly stacked in a mixture of HCP and FCC ordering of the layers. More exotic crystalline phases are possible by using charged colloids, or suspensions with colloids of multiple sizes. In addition to their esthetic properties, colloidal crystals may be useful for photonic materials [20]. For such materials, the requirement is a crystalline structure of particles with different indices of refraction from the surrounding medium; also, the lattice constant needs to be similar to the wavelength of light. Colloidal crystals fulfill both these properties, and their spontaneous self-assembly may provide a useful way to build photonic materials. We will review some recent studies of colloidal crystals; for older work, see Ref. 19.

The study of the nucleation and growth of crystals is one of the long-standing problems of solid state physics which has been recently addressed by colloidal experiments [21-23]. By taking three-dimensional images using confocal microscopy [21,22] or conventional techniques [23], critical crystal nuclei can be observed and characterized. By studying the tendency of crystalline regions of various sizes to shrink (if they are too small) or grow (if they are larger than the critical size), the critical size can be identified as O(100) particles [21]. The statistics of the smallest clusters can be related to their surface tension. Structurally, the nuclei are composed of hexagonally packed layers, in agreement with the ultimate ordering of the fully-grown crystal. The stacking between hexagonal layers slightly favors FCC packing [21,24]. These colloidal experiments are the first experiments of any sort to be able to directly study the crystal nucleation process.

The somewhat randomly stacked hexagonal layers of typical colloidal crystals are undesirable for photonic materials, and so recent work has tried to find new ways of nucleating and growing more ordered crystals. In particular, a pure FCC crystal would be useful. By finding a way to nucleate the (100) plane of an FCC crystal lattice, it may be possible to align the hexagonal layers and preserve the FCC symmetry. One possibility is to find ways to template a (100) plane onto the microscope chamber [25, 26]. The results depend on the properties of the template and the annealing process, and it appears that reasonably large FCC crystals can be grown using these methods (~30 layers deep) and having only a few defects [26]. The templating techniques are generally passive, and rely on favorable kinetics to grow the ordered crystals. An active technique is to use laser tweezers to forcibly arrange the colloidal particles, which shows great promise for creating perfect crystals [27].

*2.2 Colloidal glasses*

Colloidal crystals are relatively weak, and by gentle stirring or shaking they are easily shear-melted. In such a state, the colloidal suspension is essentially a 'super-cooled fluid': the colloidal particles are microscopically disordered, but will eventually reform into a crystal. As the concentration of a colloidal 'fluid' is increased, the motion of each colloidal particle becomes more and more restricted due to the confining effect of the other particles. Above a critical concentration (~58% for hard sphere colloids), the particles can no longer move freely through the sample, and no longer form the stable crystal state even after very long times [17*]. Such a system is considered a colloidal glass. More generally, the glassy state could be thought of as a colloidal suspension with the self-diffusion constant for particles being zero. This definition thus encompasses polydisperse colloidal suspensions that may never form crystals.

By studying dense colloidal suspensions near the glassy volume fraction, new insights into the glass transition are gained. In particular, microscopy allows direct study of dynamical heterogeneities in such systems. In a dense colloidal system, particles are confined by their neighbors, and thus large displacements occur when groups of neighboring particles all rearrange their positions simultaneously [8,9,28]. The number of particles involved in such rearrangements increases as the glass transition is approached. The difficulty of rearranging many particles simultaneously may explain why the diffusion constant for particle motion would decrease dramatically near the glass transition. Not only are these motions of particles spatially heterogeneous, they are also temporally heterogeneous, occurring in local events involving only a small fraction of the particles [8,9,28].

The gelation transition may be directly related to the colloidal glass transition, and gels are important soft materials of intrinsic interest [29]. Colloidal gels can be formed by several methods. For charge-stabilized colloids, adding in salt screens the charges, allowing the particles to approach close enough for the attractive van der Waals force to cause them to aggregate. For hard sphere colloids, a more interesting method is to add in a smaller species of particles. The small particles exert an isotropic osmotic pressure around the larger particles. Two large particles that are close to each other exclude small particles from between them, and thus feel an unbalanced osmotic pressure pushing them together [13*]. This 'depletion force' depends on the size ratio, and the concentration of the small species (which in practice may be either particles, or small polymers), and thus this attractive force can be carefully varied to study the gelation transition. Depending on the details of the depletion attraction, crystals can be found, or either diffusion-limited or reaction-limited clusters [30]. Future work should clarify connections between the glass transition of purely repulsive particles, and the gelation of attractive colloidal particles.

*2.3 Phase transitions in two dimensions*

Two dimensional systems are both simpler theoretically, and also easier to view experimentally as the restrictions on the indices of refraction of the particles and solvent are gone. One technique to produce a 2D colloidal system is to confine colloids in thin cells to prevent motion in the third dimension [15,16,31-34]. A second possibility is studying particles at an air-water interface; surface tension prevents out of plane motion [35*,36]. Another approach has used light pressure from a focused laser to push charged particles toward a charged glass wall, also effectively constraining their out of plane motion [37,38]. These approaches have been used to study the phase changes in 2D systems, and the dynamics of the colloidal particles for given phases.

A dense two-dimensional system tends to arrange into hexagonal order.  Unlike 3D crystals, these 2D crystals do not have long-range order; instead, the pair correlation function $g(r)$ decays to zero at large separations $r$.  However, the orientational order of the hexagonal lattice still is long-ranged.  The melting of these crystals is qualitatively different than the melting of three-dimensional crystals; in particular, dislocations and disclinations are important and behave differently than in three-dimensional crystals [35*,39].  In some cases, as the crystal melts, an intermediate hexatic phase is formed.  For the crystal phase, $g(r)$ decays algebraically, while for the hexatic phase, $g(r)$ decays exponentially; both phases still have short-range hexagonal order [35*,39].  As the system further melts, the hexatic phase is replaced with a liquid phase, and the hexagonal orientational order is lost.  Whether or not these distinct phases occur in experiments appears to depend on the details of the colloidal interaction.  Several experiments have seen all three phases [33,35*,36].  Another study did not find the hexatic phase, most likely as the particles had only extremely short-range repulsive interactions [31], in contrast with the previous experiments, where particles had longer-range (usually repulsive) interactions.  These experiments have provided important insight into interpreting a variety of computer simulations of 2D systems [31,35*].

An intriguing twist on these results is to study the formation of crystals in the presence of an external periodic potential, which would help create long-range order.  Similar to laser tweezers, two parallel coherent laser beams can produce interference fringes that are a periodic potential that influences colloidal particles [37].  As the external potential gradually increases, the system goes from being a modulated liquid, to a crystalline state, and then (at still higher potential) back to a modulated liquid again [37,38].  These two phase transitions are known as laser-induced freezing and laser-induced melting.  The melting may be due to a restriction on the particle motion transverse to the potential, which would reduce the effective dimensionality of the system and enhance fluctuations; the details depend on the overall particle concentration as well as the intensity of the lasers [38].

The dynamics of particles in 2D systems are also easy to study.  Dynamical heterogeneities, discussed previously in 3D liquid systems, are also found in 2D liquids [32,34].  In the crystal phases, motion occurs by defects forming and diffusing [39], and the defect motion has some surprising traits, including a temporary memory effect where defects hop between several locations for short periods [15,16].  Now that the phase diagram of 2D colloidal suspensions has been experimentally described, future work promises to relate the dynamics of particle motion directly to the phase transitions [36].

**3. Measuring interactions between colloidal particles**

*3.1 Charge-stabilized colloids*

One of the most popular colloidal systems to study is polystyrene sulfate particles in water.  Such particles are stabilized against aggregation by slight surface charges, which repel particles from each other.  The surface charges are due to the dissociation of the sulfates, and also are frequently enhanced by adsorbed surfactants.  In the water are counter-ions, which screen these repulsive interactions at large distances; this is controllable by the salt concentration within the suspension, and for very long-range repulsive interactions it is desired to have extremely pure water [40].  The presence of the counter-ions complicates the description of the inter-particle forces, especially in cases when many colloidal particles are present.  Several careful experiments have measured these inter-particle interactions.

A simple method to study interacting particles is to use laser tweezers. Two particles can be seized and moved close together by the tweezers. After releasing them, their motion is due to Brownian motion and any interactions between them [41*]. An alternate method is to trap both particles in a 'line tweezer', a laser tweezer that is rapidly scanned and which effectively confines the two particles along a line, but otherwise allows them to interact freely [13*]. In such a system again the particles' motions are due to Brownian motion and their interactions, and their confinement to the 1D line makes their motion particularly easy to follow using the particle tracking techniques discussed in Sec. 1.

The earlier experiments using these methods confirmed that the interaction between isolated pairs of charged colloidal spheres is well-described by the Derjaguin-Landau-Verwey-Overbeek theory of colloidal interactions [41*], although these results are somewhat controversial [22,42-44]. Later experiments have studied the interactions of charged particles in confined geometries, or near a charged wall [2,45-48]. In some of these experiments, particles near charged walls apparently are attracted to each other, despite their charges [45-47]. Hydrodynamic interactions between the particles and the wall appear to quantitatively explain the attractions induced by the presence of one wall [46]. The case for two-wall attractions is less clear, as they have been seen in equilibrium measurements where hydrodynamic effects should vanish [49*]. In general, these techniques can be used both to determine the form of the interaction, and also details such as specific particle charges or wall surface charge densities [2].

*3.2 Other colloidal interactions*

Attractive entropic forces between colloidal particles can be controlled by adding in a smaller species of particles, or a polymer with a small size. An isolated large colloidal particle feels an isotropic osmotic pressure from the small species, but two large particles close to each other screen out the small species between them, resulting in an unbalanced osmotic pressure that pushes the two large particles together [13*]. Laser tweezer experiments using the 'line tweezer' geometry have carefully investigated the depletion force in a variety of situations. When small colloidal spheres are the small species, at high concentration (for the small species) the exact form of the depletion force can depend on the liquid structure present within the small species [13*]. For semi-dilute polymer suspensions, the range of the depletion force depends on the correlation length scale of the polymer suspension, which is a function both of the polymer itself and also the overall polymer concentration [50]. (This assumes that the polymer does not chemically interact or bind to the particle; when it does, steric interactions of the grafted polymers result in repulsive interactions, which also have been measured [51].) The intriguing case of a highly anisotropic small species was studied by using the rodlike fd bacteriophage virus and larger colloidal particles [52]. The behavior is qualitatively similar to the simpler case of small spherical particles, but the shape of the interaction potential was noticeably modified [52].

A variety of other colloidal interactions have been studied with similar methods. The purely hydrodynamic interactions between two colloidal particles were carefully studied by two groups [53,54]. These interactions are well-described by low Reynolds number hydrodynamics [53], which also describe the motion of colloidal particles near walls [3].

Magnetorhelogical suspensions are complex fluids typically composed of soft iron particles in a solvent; applied external magnetic fields cause dramatic changes to their rheological behavior. Their magnetorheological properties depend on the interactions between the iron particles when a magnetic field is applied; typically, they form chains. The ability of these chains to break or rearrange has been studied, helping to better understand the bulk properties of the suspensions [55].

## 4. Conclusions

Many of the recent microscopic studies of colloidal suspensions focused on properties of model suspensions in equilibrium, or simple nonequilibrium phases such as glasses or gels. This suggests several new types of experiments. One possibility is to study complicated suspensions such as cornstarch in water [6]; these systems are less well-defined, but are more 'realistic' for many industrial applications. Another possibility is to examine the behavior of colloidal suspensions when they are actively forced, such as when they are sheared [56], as happens in a rheometer. This will be crucial for understanding the microscopic mechanisms leading to nontrivial frequency-dependent viscoelastic properties. High-speed cameras may be useful for probing dynamics of particles at shorter time scales. Finally, Sec. 3 described experimental work quantifying interactions between colloidal particles in simple cases. This work is a valuable foundation for designing novel materials out of simple colloidal building blocks. It is likely that such research directions will uncover new phenomena when looking at collections of different types of particles, or when several competing inter-particle interactions are present.

## Acknowledgements
We thank Gianguido Cianci, John Crocker, and Dan Young for a careful reading of the manuscript. Acknowledgement is made to the Donors of The Petroleum Research Fund, administered by the American Chemical Society, for partial support of this work. This research is also supported in part by the University Research Committee of Emory University.